\documentstyle[epsf,prb,aps,floats]{revtex}

\begin{document} 
 \draft
\wideabs{ 

\title{Mott transitions in correlated electron systems with orbital degrees
       of freedom} 

\author{Y. \=Ono,$^1$ M. Potthoff,$^{2,}$ \cite{ad} and R. Bulla$^3$} 
\address{$^1$Department of Physics, Nagoya University, Furo-cho, Chikusa-ku, Nagoya 464-8602, Japan\\
$^2$Lehrstuhl Festk\"orpertheorie, Institut f\"ur Physik, Humboldt-Universit\"at zu Berlin, D-10115 Berlin, Germany\\
$^3$Theoretische Physik III, Elektronische Korrelationen und Magnetismus, Institut f\"ur Physik, Universit\"at Augsburg,  D-86135 Augsburg, Germany}

\maketitle

\begin{abstract}
Mott metal-insulator transitions in an $M$-fold orbitally degenerate Hubbard model are studied by means of a generalization of the linearized dynamical mean-field theory. The method allows for an efficient and reliable determination of the critical interaction $U_{\rm c}$ for any integer filling $n$ and different $M$ at zero temperature. For half-filling a linear dependence of $U_{\rm c}$ on $M$ is found. Inclusion of the (full) Hund's rule exchange $J$ results in a strong reduction of $U_{\rm c}$. The transition turns out to change qualitatively from continuous for $J=0$ to discontinuous for any finite $J$. This result
is confirmed using the exact-diagonalization method for $M=2$.
\end{abstract} 
 

}

\section{Introduction}

The Mott metal-insulator transition (MIT) has been the subject of numerous experimental and theoretical investigations over the past decades \cite{Mott,Imada,FG}. As a transition from a paramagnetic metal to a paramagnetic insulator that is driven by the competition between the conduction electrons' kinetic energy and their mutual Coulomb repulsion, the Mott MIT represents a prime example for a quantum phase transition. Realizations of the Mott transition can be found in different 3d transition-metal oxides \cite{Mott,Imada} as well as in alkali-doped Fullerides \cite{Gunnarson,Koch99}, for example. Its description is a challenging task for many-body theory.

The minimal model to describe the Mott transition is the Hubbard model \cite{HM} which in its simplest form contains the local Coulomb interaction between the electrons in a conduction band formed by a {\em single} orbital only.
The investigation of such a single-band model might be justified in some cases,
in particular as an effective model for low-energy properties. 
A consistent description of the experimental observations for the above-mentioned systems, however, certainly requires the use of more 
realistic models which include orbital degrees of freedom, possible crystal field splittings, etc.

Theoretical advances in the past decade -- mainly due to the development of the dynamical mean-field theory (DMFT) \cite{MV,Georges} -- have led to an increased understanding of multi-orbital Hubbard models.
Recent investigations have either concentrated on fundamental questions such as the nature of the Mott MIT as a function of orbital degeneracy \cite{Lu94,Bue97,Fre97,Rozenberg97,Han98,Kot99,Florens,Koga}
or, within the so-called LDA+DMFT approach \cite{LDAplusDMFT}, on a realistic 
description of the transition. 

Nevertheless, there are still no reliable DMFT results for the zero-temperature ($T=0$) Mott transition in general multi-band models:
Numerical methods to solve the DMFT equations are either restricted to fairly large temperatures (quantum Monte-Carlo approach, QMC) or have not yet been extended to the multi-band case (numerical renormalization-group method, NRG).
Additionally, due to the sign problem within the QMC approach only a simplified exchange part of the Coulomb interaction can be taken into account \cite{Han98,V2O3}.
The exact-diagonalization method (ED) is restricted to rather small system sizes, and data are available for $M \le 2$ orbitals only \cite{Koga}.
The $M$ dependence of the critical interaction strength $U_{\rm c}$ for 
different integer fillings $n$ is particularly interesting.
Detailed results for $U_{\rm c}$ as a function of $M$ and $n$ are available
from the Gutzwiller approximation and from a slave-boson approach \cite{Lu94,Fre97}. 
Numerical DMFT-QMC calculations have been performed for $M \le 3$ \cite{Rozenberg97,Han98}.
Remarkably, in the limit of large orbital degeneracy $M \to \infty$ an analytical treatment of the DMFT becomes possible for the MIT \cite{Florens}. A scaling $U_{\rm c} = U_{\rm c,2} \propto M$ for the actual transition is found
while $U_{\rm c,1} \propto \sqrt{M}$ is obtained for the critical interaction where the insulating solution breaks down \cite{Florens}. This is consistent with the linear dependence for large $M$ found in Refs.\ \onlinecite{Lu94,Fre97} and with the square-root dependence reported in Ref.\ \onlinecite{Koch99}.
However, there is still a need for a DMFT method which works at $T=0$ and which allows for an efficient and reliable determination of $U_{\rm c}$ for arbitrary $n$ and $M$ and for a model including the full exchange interaction.

Here we present and apply a proper multi-band generalization of the so-called linearized DMFT (L-DMFT) \cite{BP00} which is a simple but rather successful technique that originally was developed for the critical regime of the Mott MIT in the single-band model.
Within the L-DMFT the lattice problem is mapped onto an Anderson impurity model with a single bath site only by considering a simplified self-consistency condition just at the critical point.
The approach can be considered to be the simplest non-trivial variant of the more systematic projective self-consistent method (PSCM) by Moeller et al \cite{MSK+95}.
It allows for extremely fast numerical calculations or even analytical results to estimate the critical parameters.
Extensions have been considered for the (Mott-Hubbard or charge-transfer) MITs in a $d$-$p$ model \cite{OBH01}, for a quantum critical point in the periodic Anderson model \cite{HB00}, and for the $M=2$ degenerate Hubbard model at half-filling and $J=0$ \cite{Koga}.
Furthermore, a general two-site DMFT \cite{Pot01} has been developed which reduces to the L-DMFT at the MIT but is not restricted to the critical point.
An application to multi-band systems is likewise possible but has not yet been considered.

As concerns effective single-band models, the linearized DMFT has been 
tested extensively by comparing with numerical results from the full DMFT.
For the standard Hubbard model \cite{BP00,OBHP01} but also for different 
thin-film and semi-infinite surface geometries \cite{PN99ad} as well
as for the $d$-$p$ model \cite{OO0102} a remarkable agreement has been 
found:
Detailed trends of $U_{\rm c}$ as a function of electronic model parameters and as a function of parameters characterizing the lattice geometry are predicted reliably.
For quantitative estimates errors of the order of a few percent have to be tolerated.

As it is demonstrated here, a straightforward extension of the method to multi-band systems is possible. We expect the L-DMFT to be a fast but reliable tool in this case as well since the effective impurity model includes the 
complete atomic part and the self-consistently determined bath with 
the complete degeneracy (although restricted to a single site only).
%Within the context of the PSCM, this is again equivalent to the 
%lowest-order of a systematic approach. 
Furthermore, the reliability of the L-DMFT for the multi-band case is 
checked by comparing with different fully numerical DMFT techniques 
whenever results are available. All DMFT calculations discussed
in this paper were performed on a
Bethe lattice with $t^2 = 1$, corresponding to bandwidth of 4.

\section{Model and linearized DMFT}

We consider a multi-band Hubbard-type model $H = H_0 + H_1$ 
including an intra-orbital hopping
\begin{equation}
  H_0 = \sum_{ij\alpha\sigma} t_{ij}
  c_{i\alpha\sigma}^\dagger c_{j\alpha\sigma}
\end{equation}
and the direct and the exchange part of the on-site Coulomb 
interaction:
\begin{eqnarray}
  H_1 &=& \frac{1}{2} 
  \sum_{i\alpha\alpha'\sigma\sigma'} 
  U_{\alpha\alpha'} \:
  c^\dagger_{i\alpha\sigma} c^\dagger_{i\alpha'\sigma'} 
  c_{i\alpha'\sigma'} c_{i\alpha\sigma} 
  \nonumber \\
  &+& \frac{1}{2} 
  \sum_{i\alpha\alpha'\sigma\sigma'}
  J_{\alpha\alpha'} \:
  c^\dagger_{i\alpha\sigma} c^\dagger_{i\alpha'\sigma'} 
  c_{i\alpha\sigma'} c_{i\alpha'\sigma} 
  \: ,
\label{eq:interaction}
\end{eqnarray}
using standard notations. $i$ is a site index, and 
$\sigma=\uparrow,\downarrow$ refers to the spin direction.
The different orbitals labeled by $\alpha = 1,...,M$ are considered
to be equivalent, $M$ is the orbital degeneracy.
Exploiting atomic symmetries, we have
$U_{\alpha\alpha'} = (U+2J) \delta_{\alpha\alpha'} 
                     + U (1-\delta_{\alpha\alpha'})$ and
$J_{\alpha\alpha'} = J (1-\delta_{\alpha\alpha'})$ as usual
(see e.g.\ Ref.\ \onlinecite{OS84}). 
Opposed to calculations using the QMC method
(see, e.g.\ Refs.\ \onlinecite{Han98,V2O3}),
$H_1$ includes the full exchange part and thus preserves rotational 
invariance.

The DMFT assumes the self-energy to be local.
Furthermore, the self-energy as well as the Green function are diagonal 
with respect to the orbital index $\alpha$ for the model considered here.
In case of equivalent orbitals and in the absence of any spontaneous 
symmetry breaking, we thus have 
$\Sigma_{ij\alpha\beta}(\omega) = \delta_{ij}\delta_{\alpha\beta}
\Sigma(\omega)$ and likewise for the local Green function:
$G_{ii\alpha\beta}(\omega) = \delta_{\alpha\beta} G(\omega)$.
Within the framework of the DMFT, the model $H$ is mapped onto an impurity 
model $H'=H'_0 + H'_1$ where $H'_1$ is the local part of the interaction 
(\ref{eq:interaction}) at a distinguished site $i_0$ and 
\begin{eqnarray} 
  H'_0 & = & 
  \sum_{\alpha\sigma} t_0 \:
  c^\dagger_{i_0\alpha\sigma} c_{i_0\alpha\sigma} 
  + \sum_{\alpha \sigma, k=2}^{n_{\rm s}} 
  \epsilon_{k} \: 
  a^\dagger_{k\alpha\sigma} a_{k\alpha\sigma}
\nonumber \\
  & + & 
  \sum_{\alpha\sigma, k=2}^{n_{\rm s}}
  V_{k} \: 
  (c^\dagger_{i_0\alpha\sigma} a_{k\alpha\sigma} + \mbox{h.c.}) 
\label{eq:siam} 
\end{eqnarray}
with $t_0 = t_{ii}$. 
The self-energy $\Sigma(\omega)$ of $H'$ is identified with the self-energy 
of the lattice model $H$ and yields, via the lattice Dyson equation, the 
on-site Green function $G(\omega)$.
The latter determines the one-particle parameters of $H'$ or, equivalently, 
the hybridization function
$\Delta(\omega) = \sum_k V_{k}^2 / (\omega + \mu - \epsilon_{k})$ via
the DMFT self-consistency condition
$\Delta(\omega) = \omega + \mu - t_0  - \Sigma(\omega) - 1/G(\omega)$.

Within the full DMFT an infinite number of bath degrees of freedom 
$n_{\rm s} \to \infty$ are necessary to fulfill the self-consistency 
equation. 
This implies the need for further approximations to treat $H'$.
Contrary, a simple two-site model is considered here.
For $n_{\rm s} = 2$ the model (\ref{eq:siam}) can be solved exactly; however, a simplified 
self-consistency condition must be tolerated.

Our approach rests on the assumption that similar as for the single-band
model, the MIT is characterized by a vanishing weight of a low-energy 
quasi-particle resonance.
Close to the MIT it is then plausible to neglect any internal structure 
of the quasi-particle peak and to assume that there is no influence of 
high-energy features on the low-energy part of the 
excitation spectrum.
The quasi-particle peak is then approximated by a single pole at the 
Fermi energy ($\epsilon_{\rm c} \equiv \epsilon_{k=2} = \mu$) which reproduces itself in the DMFT 
self-consistency cycle.
One can proceed as for the single-band model \cite{BP00} and is finally left with the following simple algebraic self-consistency equation 
which determines the critical parameters ($V\equiv V_{k=2}$):
\begin{equation}
   V^2 = z M_2^{(0)} \: .
\label{eq:sc}
\end{equation}
Here $M_2^{(0)}$ is the second moment of the non-interacting density
of states $M^{(0)}_{2} = \int dx \, x^2 \rho^{(0)}(x)$.
Formally, Eq.\ (\ref{eq:sc}) is the same as in the single-band model 
\cite{BP00}.
However, the quasi-particle weight $z$ has to determined from the degenerate 
model (\ref{eq:siam}).
We calculate $z = 1 / (1-\Sigma'(0))$ from the spectral representation 
of $G(\omega)$ by summing the residua $\propto V^2$ in the limit $V \to 0$.
The residua are obtained by standard (Brillouin-Wigner) degenerate 
perturbation theory up to the order $V^2$.
This yields an expression 
$z=z(V, \epsilon_{\rm c}, t_0, U, J) = 
V^2 \, F(\epsilon_{\rm c} - t_0 , U, J) / M_2^{(0)} + {\cal O}(V^4)$.
Using $\epsilon_{\rm c} = \mu$ in the limit $V \to 0$ and setting $t_0=t_{ii}=0$ to fix the energy zero, we obtain from Eq.\ (\ref{eq:sc})
$F(\mu, U, J) = 1$ as a condition for the MIT.
This condition is different for different orbital degeneracy $M$ and
integer filling $n=1,...,2M-1$.

\section{Results and discussion}
 
For an integer filling $n$ and for any $U$ larger than a critical interaction
$U_{\rm c} = U_{\rm c}(n,M)$ there are two critical values for the chemical potential
$\mu_\pm(U)$ which can be obtained by solving $F(\mu , U) = 1$ for $\mu$
(the case $J=0$ is considered first).
For $U > U_{\rm c}(n,M)$ and $\mu_-(U) < \mu < \mu_+(U)$ the system is a Mott 
insulator with integer filling $n$. 
On the other hand, for $\mu < \mu_-(U)$ or $\mu > \mu_+(U)$, 
a non-integer filling is realized, and the system becomes metallic.
The chemical potential is a continuous function of the filling $n'$ in the
vicinity of $n$ for any $U < U_{\rm c}(n,M)$ but $\mu(n')$ is discontinuous at $n'=n$
with a jump from $\mu_-(U)$ to $\mu_+(U)$ for $U > U_{\rm c}(n,M)$.
$U_{\rm c}$ is determined by minimization of the function $U(\mu)$ which is 
obtained by solving $F(\mu , U) = 1$ for $U$.

\begin{figure}[b]
\epsfxsize=2.7in
\epsffile{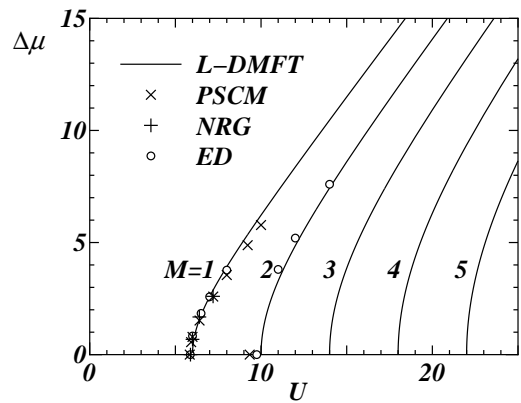}
\caption[]{
$\Delta \mu (U)$ for half-filling $n=M$ and different $M$ as obtained
from Eq.\ (\ref{eq:deltamu}). 
Results from the full DMFT using ED, NRG, and PSCM \cite{Kot99,MSK+95} are
shown for comparison.
}
\label{fig:f1}
\end{figure}

For half-filling $n = M$ and arbitrary $M$ we succeeded with a completely analytical 
calculation for $F(\mu , U)$: 
\begin{eqnarray}
F(\mu, U) / M_2^{(0)} 
  &=& \frac{1}{2} \left(\frac{M+1}{MU-\mu}+\frac{M}{\mu-(M-1)U}\right)^2  
              \nonumber \\
  &+& \frac{1}{2} \left(\frac{M}{MU-\mu}+\frac{M+1}{\mu-(M-1)U}\right)^2 .
\label{eq:F}
\end{eqnarray}
For $U<U_c$ and half-filling $n=M$ in the symmetric model, the chemical potential is fixed to $\mu=(M-\frac{1}{2})U$ due to particle-hole symmetry. Then we find $F= M_2^{(0)}(4M+2)^2/U^2$ which results in the critical interaction 
\begin{equation}
  U_{\rm c}(n=M,M) = (4M+2) \sqrt{ M_2^{(0)} } \: . 
\label{eq:uchalf}
\end{equation}
By solving $F(\mu , U) = 1$ with Eq.\ (\ref{eq:F}) for $U>U_c$ we obtain the jump in the chemical potential 
\begin{equation}
  \Delta \mu \equiv \mu_+ - \mu_- = 
  U \sqrt{B - \sqrt{B^2 -1 + \frac{U_{\rm c}^2}{U^2} }}
\label{eq:deltamu}
\end{equation}
where $B \equiv 1 + 1 / (2(2M+1)^2 (U/U_{\rm c})^2)$. 
The result for $\Delta\mu(U)$ is shown in Fig.\ \ref{fig:f1} for $M=1,...,5$. 
Here and hereafter we set $M_2^{(0)}=1$. 
For fixed $M$ and $U \to \infty$, the jump in the chemical potential 
increases linearly with $U$. 
For $U \to U_{\rm c}$,  $\Delta\mu(U)$ shows a square-root 
behaviour.

%*******************************************************************************
%*******************************************************************************

%*******************************************************************************
\begin{figure}[t]
\epsfxsize=2.7in
\epsffile{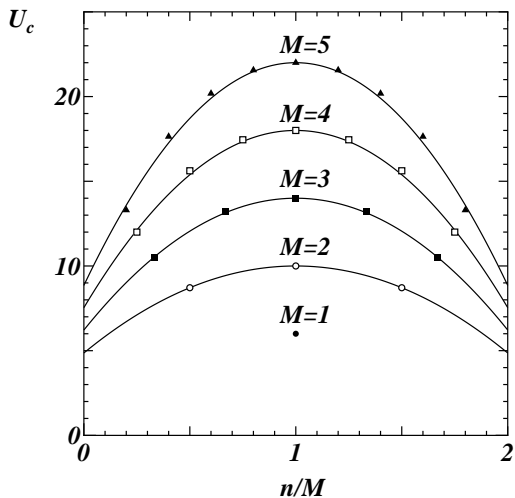}
\caption{
Critical interaction $U_{\rm c}$ for different fillings $n=1,2,...,2M-1$ and
different $M$ as a function of $n/M$.
Symbols: L-DMFT results. 
Lines: simple fit function $U_{\rm c}(n,M) = U_{\rm c}(n=M,M) - c (1-(n/M))^2$ and 
$c=-0.217 + 2.671 M$.
}
\label{fig:f2}
\end{figure}
%*******************************************************************************

Qualitatively and even quantitatively, this agrees well with results from numerical methods to solve the full DMFT equations. In particular, we compare with results from ED calculations which we have performed with $n_s = 6$ for $M=2$ (circles) and with data available for $T=0$ from
% (an essentially exact higher-order implementation of) 
the projective self-consistent method (crosses) \cite{Kot99}. The non-degenerate case $M=1$ has already been discussed in Ref.\ \onlinecite{OBHP01}. For $M=2$ and $n_{\rm s}=6$ the error due to finite-size effects in the ED is almost negligible. One can state that the agreement of the L-DMFT results with the ED and PSCM data is equally good for both $M=1$ and $M=2$.

The generalized L-DMFT predicts $U_{\rm c}(n=M,M)$ to depend linearly on $M$, see Eq.\ (\ref{eq:uchalf}). This agrees with previous results from the Gutzwiller method \cite{Lu94}, slave boson calculations \cite{Fre97}, the non-crossing approximation for low temperature \cite{Zoelfl}, and the projective technique for large $M$ \cite{Florens}. 
This finding can easily be understood by looking at the weight factors in the Lehmann representation of the spectral density of the two-site Anderson model. Expanding into the different configurations, one finds that the ground state is composed of the order of $M$ singlet states in the limit $V\to 0$ and that there are of the order of $M$ excited states coupling to the ground state. Hence, there are of the order of $M^2$ processes to be considered in the $V$-perturbation theory which are contributing to the spectral weight near $\omega = 0$. This implies $z \propto M^2 V^2 / U^2$ and eventually results in $U_{\rm c} \propto M$.

In Fig.\ \ref{fig:f2} the dependence of $U_{\rm c}$ on both $n$ and $M$ is shown
for all integer fillings and $M \le 5$. We find $U_{\rm c}$ to be reduced away from half-filling, consistent with previous results 
\cite{Koch99,Lu94,Rozenberg97,Han98,Kot99}. 
The fit shows an almost perfect quadratic dependence of $U_{\rm c}$ on the filling.

Let us now discuss the influence of the
Hund's rule coupling $J$ on the Mott transition at $T=0$.
Contrary to the QMC method, the full exchange part of the on-site Coulomb interaction can easily be included in the L-DMFT calculation. 
A quantitative influence, i.e.\ a reduction of $U_{\rm c}$ due to
$J$ has already been observed in Refs.\ \onlinecite{Koch99,Han98}. 
This behavior is also evident from our L-DMFT results, as shown below.

%*******************************************************************************
\begin{figure}[t]
\epsfxsize=2.7in
\epsffile{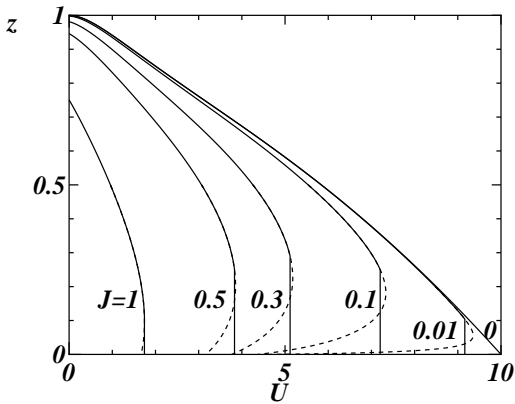}
\caption{
$U$-dependence of the quasi-particle weight $z$ for $n=M=2$ and different $J$. Dashed lines correspond to metastable solutions; vertical lines indicate the actual Mott transition at $U=U_{\rm c}$.
}
\label{fig:f3}
\end{figure}
%*******************************************************************************
%*******************************************************************************
\begin{figure}[t]
\epsfxsize=2.7in
\epsffile{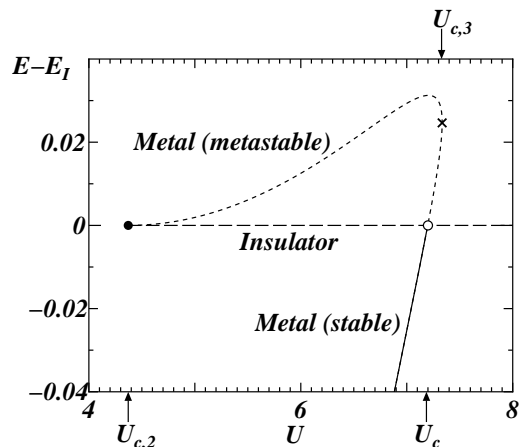}
\caption{
Energy of the two metallic solutions relative to the energy of the insulator for $J=0.1$. 
At $U_{\rm c,2}$ (closed circle) the metastable metallic state continuously merges with the 
insulating state. At $U_{\rm c,3}$ (cross) the two metallic 
states merge. The discontinuous phase transition takes place at $U_{\rm c}$ (open circle).
}
\label{fig:nf2}
\end{figure}
%*******************************************************************************

The situation is, however, more complicated: It turns out that a finite $J$ leads to a {\em qualitative} change of the nature of the $T=0$ Mott transition. This is illustrated in Fig.\ \ref{fig:f3} which shows the $U$-dependence of the quasi-particle weight $z$ for various values of $J$ and $n=M=2$. 
Note that $z$ is plotted not only close to the transition but also for $U$ far below $U_{\rm c}$ which means that 
the self-consistency relation (\ref{eq:sc}) is used for any $U$. This is reasonable as long as there is a clear separation between the low-energy (quasi-particle) and the high-energy scale (Hubbard bands). Although the approximation becomes questionable for higher $z$ and particularly in the limit $z \to 1$, the trend of $z(U)$ looks rather plausible in whole range.

In the critical regime where $z$ goes to zero 
continuously, we find $z(U)$ to decrease with {\em decreasing} $U$ which seems to be unphysical. In fact, this behavior belongs to a metastable state (dashed line) while for the true ground state (solid line) $z(U)$ is always decreasing with increasing $U$. The two metallic states with finite $z>0$ but different slope $dz/dU$ are coexisting in a finite range of interaction strengths, namely for $U$ larger than $U_{\rm c,2}$ (given by $z=0$) and for $U$ smaller than another critical value $U_{\rm c,3} > U_{\rm c,2}$ where they merge together \cite{uc1}. 

\begin{figure}[t]
\epsfxsize=2.7in
\epsffile{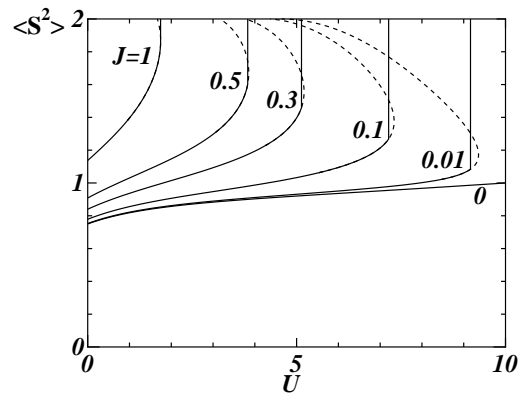}
\caption{
Local moment at the impurity site $\langle {\bf S}^2 \rangle$ as a function of $U$ for different $J$. Dashed lines: metastable solutions; vertical lines: actual Mott transition for $U=U_{\rm c}$.
}
\label{fig:nf1}
\end{figure}

The actual critical interaction strength $U_{\rm c}$ where the transition to the insulator takes place must be determined by comparing the respective energies with the energy of the insulator. This is shown in Fig.\ \ref{fig:nf2} for $J=0.1$. For any $J>0$ it is found that $U_{\rm c,2} < U_{\rm c} < U_{\rm c,3}$. At $U_{\rm c}$ the energies of the metallic and insulating states are crossing as functions of $U$. Consequently, the transition at $T=0$ is discontinuous 
with a jump in physical quantities like the double occupancy $\langle n_\uparrow n_\downarrow \rangle = dE/dU$. Equivalently, one can look at the local magnetic moment at the impurity site 
$\langle {\bf S}^2 \rangle$ (see Fig.\ \ref{fig:nf1}). 
The local moment increases with $U$, typical for a strongly correlated normal metal. At $U_{\rm c}$ it jumps to $\langle {\bf S}^2 \rangle = 2$ for any finite $J$. Similarly, the quasi-particle weigth $z$ decreases with increasing $U$ and drops to zero at $U_{\rm c}$ from a {\em finite} value $z>0$. In Fig.\ \ref{fig:f3} this is indicated by solid vertical lines. 

%*******************************************************************************
%*******************************************************************************
%*******************************************************************************
\begin{figure}[t]
\epsfxsize=2.7in
\epsffile{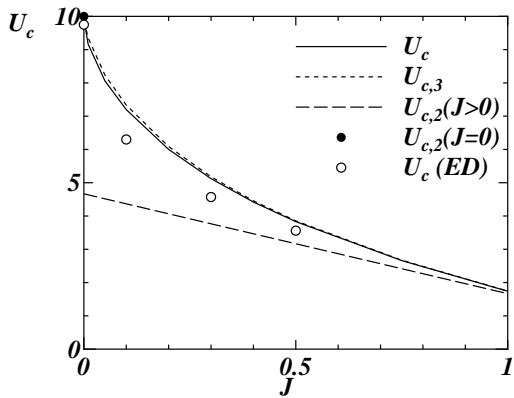}
\caption{
($U$,$J$) phase diagram for $n=M=2$.
$U_{\rm c}$: critical interaction for the MIT.
$U_{\rm c,3}$: maximum $U$ up to which a metallic solution is found.
$U_{\rm c,2}$: critical $U$ where $z \to 0$
($U_{\rm c,2} = 14/3 - 3J$ for $J>0$ and $M_2^{(0)} = 1$).
}
\label{fig:f4}
\end{figure}
%*******************************************************************************

The phase diagram in the ($U$,$J$)-plane is shown in Fig.\ \ref{fig:f4}. 
The critical interaction for the actual transition is plotted together with $U_{\rm c,2}$ and $U_{\rm c,3}$. 
$U_{\rm c}$ is found to be very close to $U_{\rm c,3}$ for all $J$-values.
Note that already a moderate value of $J \approx 0.3$ leads to a drastic reduction of $U_{\rm c}$ from $U_{\rm c}=10$ to $U_{\rm c} \approx 5$.

%*******************************************************************************
\begin{figure}[t]
\epsfxsize=3.0in
\epsffile{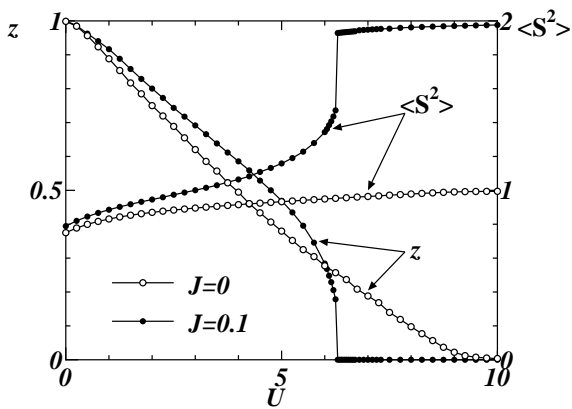}
\caption{
DMFT-ED results ($n_{\rm s}=6$) for the $U$-dependence of the quasi-particle weight (left scale) and the local moment (right scale). $J=0$ and $J=0.1$ for $n=M=2$.
}
\label{fig:nf3}
\end{figure}
%*******************************************************************************

Interestingly, similar results have been obtained beforehand by a rather different method: Applying a generalized Gutzwiller approximation to a degenerate two-band model, B\"unemann et al.\ \cite{Bue97} found the Mott transition to be discontinuous for any finite $J>0$ and to be continuous for $J=0$ only. While the Gutzwiller method (for $J=0$) significantly overestimates the critical interaction strength as compared to the L-DMFT, the reduction of $U_{\rm c}$ by a finite $J$ is stronger. Common to both methods, however, is the qualitative change of the Mott transition for a finite $J$. 

We have additionally checked the nature of the Mott transition 
by a fully numerical evaluation of the DMFT equations using the ED 
method with $n_{\rm s}=6$ sites for $n=M=2$. 
The results for $J=0$ and $J=0.1$ are shown in Fig.\ \ref{fig:nf3}. 
The critical interaction $U_{\rm c}$ is found to be somewhat smaller 
as compared to the L-DMFT result as shown in Fig.\ \ref{fig:f4}. 
 More important, however, the ED calculations show the same qualitative trend: The slope of $z(U)$ approaches $-\infty$ at a critical value $U=U_{\rm c,3}$. This clearly indicates a second metastable metallic solution and a discontinuous transition and thus corroborates our results discussed above. Fig.\ \ref{fig:nf3} also shows the $U$ dependence of the local magnetic moment where the effect is even more pronounced. Note that in the insulator $\langle {\bf S}^2 \rangle$ is close but not equal to the prediction of the linearized DMFT $\langle {\bf S}^2 \rangle = 2$ (cf.\ Fig.\ \ref{fig:nf1}). This is due to residual local fluctuations which are neglected by the L-DMFT for the insulating phase.

The limit $J \to 0$ which appears to be somewhat exceptional can be analyzed easily. From Fig.\ \ref{fig:f3} it is obvious that the critical interaction $U_{\rm c,2}$ at which $z \to 0$ in the metastable phase shows a discontinuous jump from a value $U_{\rm c,2} \approx 5 < U_{\rm c,3}$ for $J > 0$ to $U_{\rm c,2} = 10 = U_{\rm c,3}$ for $J=0$. To find the physical reason for this discontinuity of $U_{\rm c,2}$, one can again look at the Lehmann representation of the spectral density for $V \to 0$. The main difference between the case $J=0$ and the case $J>0$ consists in the fact that for $J>0$ only those configurations with a triplet on the impurity site contribute to the (singlet) ground state in the limit $V \to 0$ (see Fig.\ \ref{fig:nf1}). As compared with $J=0$ this is about a 
factor of 2 configurations less. Thus, the discontinuous jump of $U_{\rm c,2}$ is caused by a strong suppression of orbital fluctuations due to a finite $J$ (a similar argument in a slightly different context has been put forward in Ref.\ \onlinecite{Koga}). On the other hand, it should be stressed that the actual critical interaction $U_{\rm c}$ remains continuous in the limit $J \to 0$ as $U_{\rm c} \to U_{\rm c,3}$. Note that to uncover this subtle distinction it has been necessary to extend the theory beyond the $z \to 0$ limit.

\section{Conclusion}

To summarize, we have presented an extension of the linearized
dynamical mean-field theory (L-DMFT) to multi-orbital Hubbard models
including the {\em full} exchange term. 
This allows a reliable calculation of the critical interaction strength 
$U_{\rm c}$ for the zero-temperature Mott transition analytically
for all $M$ at half-filling and numerically for all integer
fillings $n\ne M$ and $M \le 5$. We find a linear dependence
of $U_{\rm c}$ on the number of orbitals. Remarkably, a qualitative change of the nature of the $T=0$ Mott transition is observed for any finite $J$: 
the transition becomes discontinuous with a finite jump of the quasi-particle weight at $U_{\rm c}$ close to $U_{\rm c,3}$.

For finite temperatures one might speculate that a finite $J$ increases the
tendency for a first-order Mott-transition. This could be of experimental
relevance for the Mott-transition in transition-metal oxides.

The L-DMFT for correlated electron systems with orbital degrees of freedom 
is a handy and trustable approach with a large potential for future applications.
One might consider, for instance, the multi-band Hubbard model investigated in the context of the LDA+DMFT approach for the transition-metal oxide V$_2$O$_3$ \cite{V2O3}.
The L-DMFT can be easily extended to answer questions about the importance of the e$_{\rm g}^\sigma$-bands (neglected in Ref.\ \onlinecite{V2O3}) on the Mott transition, or the influence of the {\em full} exchange term. 
We have already observed a significant quantitative change of $U_{\rm c}$
when excluding the spin-dependent part from the exchange term in (\ref{eq:interaction}). 
This should have consequences for a realistic description of the MIT in transition-metal oxides which are worth to be studied.

\acknowledgements

We acknowledge the support of the Grant-in-Aid for Scientific Research from the Ministry of Education, Science, Sports and Culture (Y\=O) and the Deutsche Forschungsgemeinschaft through the Sonderforschungsbereich 290 (MP) and 484 (RB).


\begin{references}
\bibitem[*]{ad} Present address: Institut f\"ur Theoretische Physik und Astrophysik, Universit\"at W\"urzburg, Am Hubland, 97074 W\"urzburg, Germany

\bibitem{Mott} N.~F.~Mott, Proc.~Phys.~Soc.~London~A~{\bf 62}, 
416 (1949); {\sl Metal-Insulator Transitions}, 2nd ed.\ 
(Taylor and Francis, London, 1990).

\bibitem{Imada} M. Imada, A. Fujimori, and Y. Tokura,
                          Rev. Mod. Phys. {\bf 70}, 1039 (1998).

\bibitem{FG} F.~Gebhard, {\sl The Mott Metal-Insulator Transition}, 
Springer Tracts in Modern Physics Vol.~137 (Springer, Berlin, 1997).

\bibitem{Gunnarson} O. Gunnarson, Rev. Mod. Phys. {\bf 69}, 575 (1997).

\bibitem{Koch99}
  E.\ Koch, O.\ Gunnarsson, and R.\ M.\ Martin, 
  Phys.\ Rev.\ B {\bf 60}, 15714
  (1999).

\bibitem{HM} J.~Hubbard, Proc.~R.~Soc.~London A~{\bf 276}, 238 (1963);
                  M.\ C.\ Gutzwiller, Phys.~Rev.~Lett.~{\bf 10}, 59 (1963);
                 J.\ Kanamori, Prog.\ Theor.\ Phys.\ {\bf 30}, 275 (1963).

\bibitem{MV} W.~Metzner and D.~Vollhardt, 
Phys.~Rev.~Lett.~{\bf 62}, 324 (1989).

\bibitem{Georges}  A.~Georges, G.~Kotliar, W.~Krauth, and
M.~J.~Rozenberg, Rev.\ Mod.\ Phys.~{\bf 68}, 13 (1996).

\bibitem{Lu94}
  J.~P. Lu,
  Phys. Rev. B {\bf 49}, 5687 (1994).

\bibitem{Bue97}
   J. B\"unemann and W. Weber,
   Phys. Rev. B {\bf 55}, 4011 (1997);
   J. B\"unemann, W. Weber, and F. Gebhard,
   Phys. Rev. B {\bf 57}, 6896 (1998).
   
\bibitem{Fre97}
   R. Fr\'esard and G. Kotliar, 
   Phys. Rev. B {\bf 56}, 12909 (1997);
   R. Fr\'esard and M. Lamboley,
   J. Low Temp. Phys. {\bf 126}, 1091 (2002).

\bibitem{Rozenberg97}
  M.\ J.\ Rozenberg, 
  Phys.\ Rev.\ B {\bf 55}, R4855 (1997).

\bibitem{Han98}
   J.~E. Han, M. Jarrell, and D.~L. Cox,
  Phys. Rev. B {\bf 58}, 4199 (1998). 

\bibitem{Kot99}
  G. Kotliar and H. Kajueter,
  Phys.\ Rev.\ B {\bf 54}, 14221
  (1999).

\bibitem{Florens}
    S. Florens, A. Georges, G. Kotliar, and O. Parcollet,
    Phys. Rev. B {\bf 66}, 205102 (2002).

\bibitem{Koga}
    A. Koga, Y. Imai, and N. Kawakami,
    Phys. Rev. B {\bf 66}, 165107 (2002). 

\bibitem{LDAplusDMFT}
     for a review see: 
     K. Held, I.~A. Nekrasov, G. Keller, V. Eyert, N. Bl\"umer, A.~K. McMahan, R.~T. Scalettar, T. Pruschke, V.~I. Anisimov, and D. Vollhardt,
%     in {\it 
%      Quantum Simulations of Complex Many-Body Systems: From Theory to Algorithms},
%      eds. J. Grotendorst, D. Marx and A. Muramatsu,
%      NIC Series Vol. 10 (NIC Directors, Forschungszentrum J\"ulich, 2002), 
%      p. 175; 
       cond-mat/0112079. 

\bibitem{V2O3}   K. Held, G. Keller, V. Eyert, D. Vollhardt, and V.~I. Anisimov, 
       Phys. Rev. Lett. {\bf 86}, 5345 (2001). 

\bibitem{BP00}
     R. Bulla and M. Potthoff,
        Eur. Phys. J. B {\bf 13}, 257 (2000).

\bibitem{MSK+95} G. Moeller, Q. Si,  G. Kotliar, M.
               Rozenberg, and D.~S. Fisher,
              Phys. Rev. Lett. {\bf 74}, 2082 (1995); 
              see also: G. Moeller, PhD-thesis,
               Rutgers (1994).

\bibitem{OBH01}  Y. \=Ono, R. Bulla, and A.~C. Hewson,
     Eur. Phys. J. B {\bf 19}, 375 (2001).

\bibitem{HB00} K. Held and R. Bulla, Eur. Phys. J. B {\bf 17}, 7 (2000).

\bibitem{Pot01} M. Potthoff, Phys. Rev. B {\bf 64}, 165114 (2001).

\bibitem{OBHP01}
       Y. \=Ono, R. Bulla, A.~C. Hewson, and M. Potthoff,
       Eur. Phys. J. B {\bf  22}, 283 (2001).

\bibitem{PN99ad} M. Potthoff and W. Nolting,
              Eur. Phys. J. B {\bf  8},  555 (1999);
              Phys. Rev. B {\bf 60}, 7834 (1999).


\bibitem{OO0102} Y. Ohashi and Y. \=Ono,
             J. Phys. Soc. Japan {\bf 70}, 2989 (2001);
            {\it ibid.} {\bf 71}  , 217  (2002) Suppl..

\bibitem{OS84} A.~M. Ole\'s{} and G. Stollhoff,
             Phys. Rev. B {\bf 29}, 314 (1984).

\bibitem{Zoelfl} M. Z\"olfl, PhD-thesis, University of Regensburg (2001).

\bibitem{uc1} 
Taking into account additionally the insulating ($z=0$) solution for 
$U > U_{\rm c,1}$ ($U_{\rm c,1} \le U_{\rm c,2}$), this amounts to {\em three} coexisting solutions of the mean-field equations in the above interval. 

\end{references}
\end{document}